\documentstyle[A4,12pt]{article}
\title{Exact solutions of the associated Camassa-Holm equation}

\author{A.N.W. Hone \\ 
\normalsize 
\em Dipartimento di Fisica `E. Amaldi', \\ 
\normalsize
\em Universita degli Studi `Roma Tre', \\ 
\normalsize
\em Roma, Italia. \\
\normalsize
e-mail: hone@amaldi.fis.uniroma3.it} 

\begin{document} 
\renewcommand{\theequation}{\arabic{section}.\arabic{equation}} 
\maketitle 

\begin{abstract} 

Recently the associated Camassa-Holm (ACH) equation, related to the 
Fuchssteiner-Fokas-Camassa-Holm (FFCH) equation by a hodograph transformation, 
was introduced  by Schiff, who derived B\"{a}cklund transformations by a loop 
group technique and used these to obtain some simple soliton and rational solutions. 
We show how the ACH equation is related to Schr\"{o}dinger operators and the KdV 
hierarchy, and use this connection to obtain exact solutions (rational and 
N-soliton solutions). More generally, we show that  solutions of ACH  on a constant non-zero 
background can 
be obtained directly from the tau-functions of 
known solutions of the KdV hierarchy on a zero background. 
We also present exact solutions given by a particular case of the third Painlev\'{e} 
transcendent.  

\end{abstract} 

\section{Introduction}

A great deal of interest has been generated by the Fuchssteiner-Fokas-Camassa-Holm (FFCH) 
equation, 
\begin{equation} 
u_{T}=2f_{X}u+fu_{X}, \qquad u=\frac{1}{2}f_{XX}-2f 
\label{eq:ch} 
\end{equation} 
(we are using the slightly non-standard choice 
of coefficients taken in \cite{schiff}), which originally 
appeared in the work of Fuchssteiner and Fokas \cite{ff}, but was later derived  
as an 
equation for shallow water waves by Camassa and Holm \cite{ch}.    
Of particular importance was the discovery \cite{ch} 
that (\ref{eq:ch}) admits peaked solitons or  ``peakons", described in terms of solutions 
of an associated integrable finite-dimensional 
dynamical system which has subsequently been related to the Toda lattice \cite{ragn}.  
Although the FFCH equation is integrable, and  
has been shown \cite{fuch} to be related by a hodograph transformation to the   
first negative flow of the KdV hierarchy (also known as the AKNS equation \cite{akns}),  
\begin{equation} 
R(U) \, U_{t}=0, \qquad
R(U)=\partial_{x}^{2}+4U+2U_{x}\partial_{x}^{-1}, 
\label{eq:invkdv}
\end{equation} 
it has many non-standard features (for instance, it possesses only the weak Painlev\'{e} 
property \cite{pick}) and there is still much to be understood about its solutions. 
We refer the reader to \cite{schiff} for a more complete list of references 
concerning (\ref{eq:ch}).   

Inspired by \cite{fuch}, Schiff introduced the associated Camassa-Holm 
(ACH) equation \cite{schiff} 
 \begin{equation} 
 p_{t}=p^{2}f_{x}, \qquad f=\frac{p}{4}(\log [p])_{xt}-\frac{p^{2}}{2}, 
 \label{eq:ach} 
 \end{equation}    
which (for positive $u$) has a one to one correspondence with solutions of the FFCH 
equation (\ref{eq:ch}) given by 
\begin{equation}  
p=\sqrt{u}, \qquad dx=p\, dX + pf \, dT, \qquad dt=dT  
\label{eq:hod} 
\end{equation}  
(the independent variables $x,t$ of ACH are denoted $t_{0},t_{1}$ in \cite{schiff});  
a solution of ACH where $p$ has zeros corresponds to a number of solutions of 
(\ref{eq:ch}) where $u$ has fixed sign. In \cite{schiff}, a loop group interpretation 
was given for (\ref{eq:ach}), making use of the fact that it is the (zero curvature) 
compatibility condition for the linear system 
\begin{eqnarray} 
\left( \begin{array}{c} \psi_{1} \\ \psi_{2} \end{array} \right) _{x}
& = & \left( \begin{array}{cc}  
$$ 0$$ & $$1/p $$ \\ 
$$ p/\lambda + 1/p $$ & $$ 0 $$ \end{array} \right)  
\left( \begin{array}{c} \psi_{1} \\ \psi_{2} \end{array} \right),     
\label{eq:zero} 
\end{eqnarray}  
\begin{eqnarray}    
\left( \begin{array}{c} \psi_{1} \\ \psi_{2} \end{array} \right) _{t}
& = & \left( \begin{array}{cc}  
$$ -p_{t}/2p $$ & $$\lambda $$ \\ 
$$ \lambda -2f $$ & $$ p_{t}/2p $$ \end{array} \right)   
\left( \begin{array}{c} \psi_{1} \\ \psi_{2} \end{array} \right),   
\label{eq:zero2} 
\end{eqnarray} 
and it was indicated that this ACH equation is part of an 
integrable hierarchy of zero curvature equations. Automorphisms of the loop group were used  
 to derive two B\"{a}cklund transformations (BTs) for (\ref{eq:ach}), and these were applied to  
the constant background solution $p=h$ to obtain some interesting solutions, 
in particular a simple rational solution, one- and two-soliton 
solutions (all on the same constant background) 
as well as a superposition formula. By applying the hodograph transformation (\ref{eq:hod}) 
apparently novel solutions of the FFCH equation (\ref{eq:ch}) were generated (although 
it seems that in general the quadratures involved cannot be evaluated explicitly).    
  
In the following we show that, since the hodograph 
transformation is essentially the same as in \cite{fuch}, 
the ACH equation 
is also related to the inverse KdV equation (\ref{eq:invkdv}) and has 
a simple Lax pair of which one part is just a (time-independent) Schr\"{o}dinger 
equation. Thus we see that the 
first BT presented in \cite{schiff}, and the 
corresponding solutions, can be derived by considering 
the usual  Crum transformation for Schr\"{o}dinger operators \cite{admo}
 and further show that a solution of ACH on a constant non-zero background may be 
 obtained from the tau-function of 
 a corresponding solution of the KdV hierarchy on a zero background. 
Rational solutions of ACH are given in terms of  the Adler-Moser 
 polynomials \cite{admo} for KdV, while the N-soliton on constant background may be obtained 
 immediately from   the standard formulae for KdV N-soliton solutions. 

In the next section we present some simple properties and solutions of (\ref{eq:ach}), 
as well as summarizing    
the BTs  introduced in \cite{schiff} and showing how they are related to the Schr\"{o}dinger 
equation. Section three 
contains a demonstration that solutions of (\ref{eq:ach}) on a constant background 
can be obtained from the tau-functions of solutions 
of the KdV hierarchy vanishing at infinity.  The fourth section contains a brief summary 
of the exact solutions of (\ref{eq:ach}) we have obtained, in particular 
rational and soliton solutions (which are derived as a corollary of the result in 
section three), as well as a conjectured form of elliptic solutions, and solutions 
in terms of the third Painlev\'{e} transcendent (PIII). 

\section{Basic  properties of the ACH equation} 

\setcounter{equation}{0} 

\subsection{Simple solutions, Lax pair and tau-function} 

In order to explore the properties of the ACH equation (\ref{eq:ach}), we find it is 
convenient to rewrite it in two different ways. For analyzing reductions of the equation, it is 
useful to write it in the form 
\begin{equation} 
 4[p^{-1}]_{t}+\left[p^{-1}(pp_{xt}-p_{x}p_{t}-2p^{3})\right]_{x}=0. 
 \label{eq:bach} 
\end{equation} 
The simplest solutions come from the travelling wave reduction (also considered in 
\cite{schiff}), and it is easy to see that the general solution of this type is 
given in terms of the Weierstrass $\wp$-function \cite{whit}, 
\begin{equation} 
p(x-ct)=-c\wp(x-ct)+k   
\label{eq:ellip} 
\end{equation} 
($k$ is constant); the special case $c=0$ yields the solution $p=const$. 
Degenerations of the $\wp$-function  give 
the one-pole rational solution and one-soliton solution 
(both on a constant background $h$), 
respectively 
\begin{equation}
p=h-\frac{h^{3}}{(x-h^{3}t)^{2}}, \qquad 
p=\left(h+\frac{\lambda}{h}\right)\tanh^{2}\left(\sqrt{\frac{1}{h^{2}}+\frac{1}{\lambda}} 
(x+\lambda h t )\right) -\frac{\lambda}{h} .   \label{eq:solu} 
\end{equation} 

In the form (\ref{eq:bach}) it is also straightforward to obtain the scaling similarity 
reduction of the ACH equation, 
$$ 
p=(2t)^{-\frac{1}{2}}w(z), \qquad z=(2t)^{\frac{1}{2}}x, 
$$ 
where $w(z)$ satisfies the ODE 
\begin{equation} 
w''=\frac{(w')^{2}}{w}-\frac{w'}{z}+\frac{1}{z}\left(2w^{2}+\beta\right)-\frac{4}{w}; 
\label{eq:pth} 
\end{equation} 
$'$ denotes $d/dz$ and $\beta$ is an arbitrary constant. This ODE is a special case 
of the Painlev\'{e} transcendent PIII \cite{pth}, and we shall consider it again in section 4. 
For now we simply observe that for $\beta=0$  it admits the particular solution 
$w=(2z)^{\frac{1}{3}}$.  

The second way to rewrite the ACH equation is 
\begin{equation} 
U_{t}=-2p_{x}, 
\qquad U=-\frac{1}{2}\left( \frac{pp_{xx}-\frac{1}{2}p_{x}^{2}+2}{p^{2}}\right). 
\label{eq:cach} 
\end{equation} 
This is the best way to write it, as it leads directly to a connection with the KdV hierarchy. 
In fact, by writing $p=-\frac{1}{2}\partial_{x}^{-1}U_{t}$, it is possible to show directly 
that $U$ satisfies the inverse KdV equation (\ref{eq:invkdv}). This suggests that there 
should be a direct link with Schr\"{o}dinger operators, and indeed this is the case 
(I am grateful to Decio Levi and Orlando Ragnisco for pointing this out). 

Setting 
$\phi=p^{\frac{1}{2}}\psi_{1}$ (and using $\psi_{2}=p\psi_{1,x}$) 
in (\ref{eq:zero}), (\ref{eq:zero2})  
leads to the Lax pair 
\begin{equation} 
(\partial_{x}^{2}+U-1/\lambda)\phi  =  0, \label{eq:schrod} 
\end{equation} 
\begin{equation} 
\phi_{t}=\lambda(p\phi_{x}-\frac{1}{2}p_{x}\phi).  
\label{eq:time} 
\end{equation}  
The two compatibility conditions for this Lax pair (assuming that $U$ is 
as yet undefined)  are simply the first equation for $U_{t}$ in (\ref{eq:cach}) and 
$$  
(\partial_{x}^{3}+4U\partial_{x}+2U_{x})p=0,   
$$ 
and it is  an immediate consequence  that $U$ satisfies (\ref{eq:invkdv}). The latter 
 third order equation for $p$ can be integrated once to give 
\begin{equation} 
pp_{xx}-\frac{1}{2}p_{x}^{2}+2Up^{2}+g(t)=0. \label{eq:pinney} 
\end{equation} 
Equation (\ref{eq:pinney}) is known as the Ermakov-Pinney equation 
(see \cite{common} and references). The function $g(t)$ is arbitrary, but it is clear that the ACH   
equation (with the definition of $U$ as in (\ref{eq:cach})) corresponds to the particular choice 
$g=2$. Thus solutions of the ACH equation correspond to a particular class of solutions 
of the inverse KdV equation. 

We have also applied the WTC generalization of the Painlev\'{e} test \cite{wtc} 
and found that it is satisfied by the ACH equation.  Laurent expansions 
around a non-characteristic 
singularity manifold $\zeta(x,t)=0$ have leading order terms  
$p=-(\log[\zeta])_{xt}+\ldots$, and this principal balance has resonances  $-1, 2,4$. 
Instead of considering such expansions in more detail, we merely note the connection 
between the Painlev\'{e} property and Hirota's method \cite{gib}, and thus define 
the tau-function of ACH by a truncation of this expansion: 
$$  
p(x,t)=-(\log[\tau])_{xt}. 
$$ 
Note that from the ACH equation in the form (\ref{eq:cach}) we can integrate once 
to find the standard KdV formula for the potential of the Schr\"{o}dinger operator:  
$$ 
U(x,t)=2(\log[\tau])_{xx}. 
$$  
The tau-function satisfies a trilinear equation in $x,t$ 
(this is a reduction of the trilinear appearing in \cite{schiff2}).Z
We observe that for the constant background solutions 
of ACH constructed in section 4 this $\tau$ is a tau-function of the KdV hierarchy after 
rescaling by a factor $\exp[-x^{2}/(4h^{2})-hxt]$; similarly for the solutions 
given in terms of the third Painlev\'{e} transcendent,  
$\tau$ should correspond to the tau-function for PIII introduced by Okamoto 
\cite{okamoto}.  

\subsection{B\"{a}cklund transformations} 

By the use of loop group techniques, Schiff \cite{schiff} derived two BTs for 
the ACH equation, which he considered in the original form (\ref{eq:ach}). The first 
of these BTs is  
\begin{equation} 
\tilde{p}=p-s_{x}, \quad s_{x}=-(p\lambda)^{-1}s^{2}+\lambda p^{-1}+p, 
\quad  s_{t}=-s^{2}+(\log [p])_{t}s+\lambda(\lambda-2f), 
\label{eq:bt1} 
\end{equation} 
where $\lambda$ is a B\"{a}cklund parameter. A superposition principle was 
also found for this BT, leading to a formula for the 2-soliton solution of ACH. Schiff's 
second BT may be written (after some simplification) as  
\begin{equation} 
\tilde{p}=p-(\log [\chi])_{xt}, \quad (pB_{x})_{x}=B(p^{-1}+p\lambda^{-1}), 
\quad B_{t}=-\frac{1}{2}(\log [p])_{t}B+p\lambda B_{x},  
\label{eq:bt2} 
\end{equation} 
where $\chi$ is determined from the first order equations 
$$ 
\chi_{x}=p\lambda^{-1}B^{2}, \qquad \chi_{t}=\lambda(p^{2}B_{x}^{2}-B^{2}). 
$$ 

We observe that (as also noted in \cite{schiff}) the equations for $B$ in 
(\ref{eq:bt2}) follow from the linearization of the first Riccati equation in 
(\ref{eq:bt1}) via the substitution $s=p\lambda(\log [B])_{x}$; this linearization 
is just the linear problem (\ref{eq:zero}) when we identify $B=\psi_{1}$. 
It seems that the second BT actually gives nothing new, as it appears to be equivalent to 
applying (\ref{eq:bt1}) twice with the same B\"{a}cklund parameter each time. 
Thus we concentrate on the first BT, and observe that it is more convenient to 
linearize the second Riccati equation in (\ref{eq:bt1}) by the substitution 
$s=(\log [\phi])_{t}$, and then it is straightforward to show that the BT (\ref{eq:bt1}) 
is equivalent to  
$$ 
\tilde{p}=p-(\log [\phi])_{xt}, 
$$ 
where $\phi$ is a solution of the Lax pair (\ref{eq:schrod}), (\ref{eq:time}). In fact, it is 
further possible to show (by a tedious direct calculation) that under this BT the potential 
of the Schr\"{o}dinger operator becomes 
$$ 
\tilde{U}=U+2(\log [\phi])_{xx}, 
$$ 
and also that $\tilde{\phi}=\phi^{-1}$ is a solution of the same Lax pair with $U,p$ 
replaced by $\tilde{U}, \tilde{p}$. Thus it is apparent that the first BT 
(\ref{eq:bt1}) derived by Schiff is equivalent to the well known Crum transformation 
obtained by factorization of  the Schr\"{o}dinger operator \cite{admo}, which gives 
the standard Darboux-B\"{a}cklund transformation for the KdV hierarchy.       
 
Having seen the connection with the Crum transformation, it is 
clear that it should be possible to express solutions of the ACH equation in terms of 
known solutions of the KdV hierarchy obtained by this transformation. 
Before proving a more general statement in the 
next section, we briefly illustrate this idea by presenting the simplest rational solutions. 
Applying the first BT (\ref{eq:bt1}) above starting from the constant background solution 
$p_{0}=h$, and choosing the B\"{a}cklund parameter $\lambda=-h^{2}$ each time, 
we obtain solutions of the form 
$$ 
p_{k}=h- (\log [\theta_{k}])_{xt}, 
$$
where the polynomials $\theta_{k}$ are given by 
$$ 
\theta_{0}=1, \quad \theta_{1}=\tau_{1}, \quad \theta_{2}=\tau_{1}^{3}+\tau_{2}, 
\quad \theta_{3}=\tau_{1}^{6}+5\tau_{2}\tau_{1}^{3}+\tau_{3}\tau_{1}-5\tau_{2}^{2}, 
$$ 
where in terms of $x,t$ 
$$ 
\tau_{1}=\tilde{\tau}_{1}+ x-h^{3}t, \quad \tau_{2}=\tilde{\tau}_{2}+3h^{5}t,  
\quad \tau_{3}=\tilde{\tau}_{3}-45h^{7}t  
$$ 
(for $ \tilde{\tau}_{j}$ independent of $x,t$). The $\theta_{k}$ are the 
Adler-Moser polynomials \cite{admo}, which are the tau-functions of the vanishing 
rational solutions of the KdV hierarchy. The $\tau_{j}$ are the times of the hierarchy, up  
to a suitable scaling; in the next section we will choose a more canonical 
normalization for these times. 

\section{Solutions of ACH from KdV} 
\setcounter{equation}{0} 

Before stating and proving the main result, we fix our conventions by briefly 
reviewing well known properties of the KdV hierarchy and its tau-functions  
(which may be found in many places, \cite{newell} for example).  The KdV hierarchy is 
the sequence of evolution equations 
\begin{equation} 
q_{t_{2j-1}}=2(P_{j}[q])_{t_{1}} 
\label{eq:kflo}  
\end{equation} 
(for $j=1,2,3,\ldots$), which arise as the compatibility conditon for 
the Schr\"{o}dinger equation 
\begin{equation} 
(\partial_{t_{1}}^{2}+q)\phi=\mu^{2}\phi 
\label{eq:nuschro} 
\end{equation} 
with the sequence of linear problems 
  \begin{equation} 
\phi_{t_{2j+1}}=\Pi_{j}\phi_{t_{1}}-\frac{1}{2}\Pi_{j, t_{1}}\phi, 
\quad \Pi_{j}[q;\mu] :=\sum_{k=0}^{j}P_{j-k}[q]\mu^{2k}, \quad P_{0}=1 
\label{eq:linti} 
\end{equation} 
(for $j=0,1,2,\ldots$).  
The $t_{2j-1}$ are the times of the hierarchy (the odd times of the KP hierarchy 
\cite{ohta}),     
and the differential polynomials $P_{k}[q]$ are the Gelfand-Dikii polynomials \cite{gelfand}, 
which  can be defined recursively using a form of the Ermakov-Pinney equation (\ref{eq:pinney}). 
Also, in terms of the tau-function $\tau(t_{1},t_{3},t_{5},\ldots)$ of 
the hierarchy, $q$ and the $P_{j}$ are given by   
\begin{equation} 
q=2(\log[\tau])_{t_{1}t_{1}}, \qquad P_{j}=(\log[\tau])_{t_{1}t_{2j-1}}. 
\label{eq:tauks} 
\end{equation}  
This tau-function satisfies a sequence of bilinear equations \cite{ohta}, but we 
shall not make use of these here.

To make the connection with the ACH equation we simply observe that for a solution 
of ACH satisfying $p\rightarrow h$ at infinity, it is clear that $U$ as defined in 
(\ref{eq:cach}) satisfies $U\rightarrow -1/h^{2}$. Thus considering the Schr\"{o}dinger equation 
(\ref{eq:schrod}) with such a potential $U$ is equivalent to 
instead taking a Schr\"{o}dinger equation  
 (\ref{eq:nuschro}) with potential $q$ vanishing at infinity, when we identify 
 \begin{equation}
q=U+1/h^{2}, \qquad \mu=\sqrt{\frac{1}{h^{2}}+\frac{1}{\lambda}}. 
\label{eq:conn} 
\end{equation} 
This immediately suggests the following 
 
 \noindent
 {\bf Proposition. } {\it Given a tau function $\tau(t_{1},t_{3},t_{5},\ldots)$ for a 
 solution $q(t_{1},t_{3},t_{5},\ldots)$ of the KdV hierarchy satisfying $q\rightarrow 0$ as 
 $|t_{1}|\rightarrow\infty$, a corresponding solution of the ACH equation 
 (\ref{eq:cach}) on constant background $h$ 
 is given by }
 \begin{equation} 
 p=h-(\log[\tau])_{xt}, \quad t_{1}=\tilde{t}_{1}+x-h^{3}t, \quad 
 t_{2j+1}=\tilde{t}_{2j+1}-h^{2j+3}t \quad (j\geq 1)  
 \label{eq:pdeft} 
 \end{equation}  
 {\it (with $\tilde{t}_{2j+1}$ independent of $x,t$). 
 The corresponding Schr\"{o}dinger potential is given by }
 $$ 
 U=q-1/h^{2}=-1/h^{2}+2(\log[\tau])_{xx}. 
 $$ 
 
Clearly a more general statement is possible, applying to 
non-vanishing (non-constant background) solutions also, but 
for these purposes we are concerned with the explicit 
appearance of the background parameter $h$. 
The proof of the proposition is very straightforward, for using (\ref{eq:pdeft}) 
and (\ref{eq:tauks}) we see that we can write 
$$ 
p=h+\sum_{j=1}^{\infty}h^{2j+1}(\log[\tau])_{t_{1}t_{2j-1}}
=\sum_{j=0}^{\infty}P_{j}h^{2j+1}. 
$$ 
Series of this type are a standard tool for obtaining recursive formulae for the flows of 
integrable hierarchies based on Schr\"{o}dinger operators (see \cite{anton}, for instance). 
Thus the right hand equation in (\ref{eq:cach}) (the Ermakov-Pinney equation) is naturally  
rewritten as  
$$ 
pp_{t_{1}t_{1}}-\frac{1}{2}p_{t_{1}}^{2}+2(q-1/h^{2})p^{2}+2=0, 
$$ 
and by expanding in powers of $h$  
the recursion relations for the Gelfand-Dikii polynomials $P_{j}$ are obtained. Hence 
$p$ as defined above is automatically a solution of this Ermakov-Pinney equation, 
and writing everything in terms of the tau-function the ACH equation itself 
(the left hand equation in (\ref{eq:cach})) is just the tautology 
$2(\log[\tau])_{xxt}=2(\log[\tau])_{xtx}$. 

It is also fairly straightforward to show that, provided $\mu$ is 
identified as in (\ref{eq:conn}),  $\phi$ satisfying (\ref{eq:nuschro}) 
and the sequence of linear problems  
(\ref{eq:linti})  provides a solution to the ACH Lax pair (\ref{eq:schrod}), (\ref{eq:time}) 
(thus providing an alternative proof of the proposition). In order to show that 
(\ref{eq:time}) is satisfied, it is necessary to write 
$$ 
\phi_{t}=-\sum_{j=0}^{\infty}h^{2j+3}\phi_{t_{2j+1}} 
=-\sum_{j=0}^{\infty}h^{2j+3}
\left(\Pi_{j}\phi_{t_{1}}-\frac{1}{2}\Pi_{j, t_{1}} \phi\right);  
$$ 
expanding each $\Pi_{j}$ in $\mu$ and resumming by use of the 
geometric series 
$$ 
\lambda=\frac{-h^{2}}{1-\mu^{2}h^{2}}=-\sum_{k=0}^{\infty}h^{2j+2}\mu^{2j}, 
$$ 
and noting that $t_{1}$ derivatives may be replaced by $x$ derivatives, 
(\ref{eq:time}) results. 

\section{Exact solutions} 

\setcounter{equation}{0}  

\subsection{Rational and soliton solutions}  

It is now a simple matter to obtain solutions of the ACH equation from known solutions 
of KdV. For instance, the sequence of rational solutions mentioned at the end of section 2 
correspond to tau-functions $\tau^{(k)}$ which 
are most easily expressed as Wronskian determinants of odd Schur 
polynomials,  
$$ 
\tau^{(k)}=[p_{2k-1},p_{2k-3}, \ldots, p_{1}] 
$$ 
for $k=1,2,3,\ldots$. The sequence of Schur polynomials may be defined by a generating 
function, $\sum_{l=0}^{\infty}p_{l}\nu^{l}=
\exp\left[\sum_{j=1}^{\infty}t_{j} \nu^{j}\right]$ (see e.g. \cite{ohta}).  The above Wronskians 
are independent of the even times $t_{2k}$,   
and we are using the notation $[\ldots ]$ to denote the Wronskian  as in \cite{hone}. Thus 
we find the sequence of rational   
solutions 
$$ 
p^{(k)}=h-(\log[\tau^{(k)}])_{xt}, 
$$ 
where the $t_{j}$ are given in terms of $x,t$ as in (\ref{eq:pdeft}). 
This agrees with the formulae in section 2, since after rescaling the $\theta_{k}$ introduced 
by Adler and Moser \cite{admo} are the same as these $\tau^{(k)}$. 

Similarly it is well known that the tau-functions for soliton solutions  of integrable 
hierarchies can be written as Wronskian  determinants \cite{ohta, satsuma}. 
For KdV the N-soliton tau-function is built out of $N$ functions $\eta_{j}$ of 
the form 
$$ 
\eta_{j}=\exp[\xi(t_{1}, t_{3}, \ldots;\mu_{j})]+c_{j}\exp[\xi(t_{1}, t_{3}, \ldots;-\mu_{j})], 
$$  
where $\quad \xi(t_{1}, t_{3}, \ldots;\mu)=\sum_{k=1}^{\infty}t_{2k-1}\mu^{2k-1}$. 
 Using the expressions (\ref{eq:pdeft}) for the $t_{2j-1}$, 
 and defining $\lambda_{j}=-h^{2}(1-\mu_{j}^{2}h^{2})^{-1}$ (which may 
 be written as a  geometric 
 series as in the previous section), we find that we may write 
 $$ 
 \eta_{j}=\exp\left[ \sqrt{\frac{1}{h^{2}}+\frac{1}{\lambda_{j}}}\left( 
 x-h\lambda_{j}t+x_{j}\right) \right] +c_{j} 
 \exp\left[-\sqrt{\frac{1}{h^{2}}+\frac{1}{\lambda_{j}}}\left( 
 x-h\lambda_{j}t+x_{j}\right) \right],  
 $$ 
 for $x_{j}, c_{j}$ constants (compare with the 1-soliton formula (\ref{eq:solu})). Then 
 the N-soliton solution of the ACH equation may be written as 
 $$ 
 p(N)=h-(\log[W(N)])_{xt}, \qquad W(N)=[\eta_{1}, \eta_{2}, \ldots, \eta_{N}], 
 $$ 
 (where all the $t_{1}$ derivatives from the KdV Wronskian formula \cite{satsuma} 
 may be replaced by $x$ derivatives). 

\subsection{Elliptic solutions} 

Given that the ACH equation admits the simple elliptic solution (\ref{eq:ellip}), 
it would be interesting to see whether more general solutions could be obtained 
from the ansatz 
$$ 
p=-\dot{x}_{j}\wp(x-x_{j})+const, 
$$ 
where the poles $x_{j}=x_{j}(t)$  depend on time, and $\dot{x}_{j}=\frac{d}{dt}x_{j}$. 
In the light of known results on KdV \cite{krich} we would expect that this ansatz should 
be correct provided that the poles move according to a constrained elliptic 
Calogero-Moser system.  

\subsection{Solutions in terms of PIII} 

As we have already seen, the ACH equation has a scaling  similarity reduction leading to 
a particular case (\ref{eq:pth}) of PIII, which (with the standard form of PIII as 
in \cite{pth}) corresponds to the choice of parameters $\alpha=2, \gamma=0, \delta=-4$ 
(and $\beta$ remaining arbitrary).  PIII has a large number of BTs, 
and some special exact solutions, which are systematically catalogued in \cite{pth}. However, 
for the choice of parameters relevant here,   the only BTs that survive are one referred to in 
\cite{pth} as transformation V, 
$$ 
\tilde{w}=\frac{zw'}{w^{2}}-\frac{\beta+2}{2w}+\frac{2z}{w^{2}}, 
$$ 
together with its inverse 
 $$ 
 w=-\frac{z\tilde{w}'}{\tilde{w}^{2}}-\frac{\beta-2}{2\tilde{w}}+\frac{2z}{\tilde{w}^{2}}.  
$$
These BTs send a solution to (\ref{eq:pth}) for parameter $\beta$ to another solution 
with parameter $\beta+4$, $\beta-4$
respectively.  

For this special choice of parameter values there is a hierarchy of special solutions 
rational in $z^{\frac{1}{3}}$, which can be obtained by applying the BTs to the special seed 
solution $w_{0}=(2z)^{\frac{1}{3}}$ for $\beta=0$.  For example, for $\beta=\pm 4$ 
(\ref{eq:pth}) admits the special solutions  
$$ 
w_{\pm 4}=
\frac{6 z^{\frac{2}{3}} \mp 2^{\frac{4}{3}} }{ 3. 2^{\frac{2}{3}}z^{\frac{1}{3}}}. 
$$ 
Other solutions in this hierarchy can be obtained from table 6 of \cite{pth} (on setting  
the parameters $\mu=2, \kappa=2^{\frac{1}{3}}$).

\section{Conclusions} 
We have shown how the ACH equation introduced by Schiff is related to the KdV 
hierarchy, and used this connection to construct a variety of exact solutions. We have also 
found solutions in terms of a particular case of the Painlev\'{e} transcendent 
PIII. By the use of the hodograph transformation (\ref{eq:hod}) 
these solutions of the ACH equation yield solutions of the FFCH equation (\ref{eq:ch}), 
and it would be interesting to study the transformed solutions. 
We have also found \cite{hone2} that similar methods apply 
to the 2+1-dimensional generalization of the 
FFCH equation introduced in \cite{cpg}. 

\section{Acknowledgements} 

It is a pleasure to thank Orlando Ragnisco, Decio Levi,
Andrew Pickering and Jeremy Schiff for useful discussions. 
I would also like to thank Sergei Manakov and others present 
at the Integrable Systems Seminar of Roma 'La Sapienza' 
for helpful comments.
I am very grateful to 
the Leverhulme Trust for giving me a Study Abroad Studentship in Rome.


\begin{thebibliography}{99}  

\bibitem{akns}M.J.~Ablowitz, D.J.~Kaup, A.C.~Newell and H.~Segur, 
Stud. Appl. Math. 8 (1974) 249-315. 

\bibitem{admo}M.~Adler and J.~Moser, Commun. Math. Phys. 61 (1978) 1-30.  

\bibitem{anton}M.~Antonowicz and A.P.~Fordy, Commun. Math. Phys. 124 (1989) 
465-486. 

\bibitem{pth}A.P.~Bassom, P.A.~Clarkson and A.E.Milne, 
Stud. Appl. Math. 98 (1997) 139-194. 

\bibitem{ragn}M.~Bruschi and O.~Ragnisco, Physica A 228 (1996) 150-159.  

\bibitem{ch}R.~Camassa and D.D.~Holm, Phys. Rev. Lett. 71  (1993) 1661-1664.  

\bibitem{cpg}P.A.~Clarkson, P.R.~Gordoa and A.~Pickering, Inverse Problems 13 
(1997) 1463-1476.  

\bibitem{common}A.K.~Common and M.~Musette, Phys. Lett. A 235 (1997) 574-580. 

\bibitem{ff}A.S.~Fokas and B.~Fuchssteiner, Physica D 4 (1981) 47-66.

\bibitem{fuch}B.~Fuchssteiner, 
Physica D 95 (1996) 229-243. 

\bibitem{gelfand}I.M.~Gelfand and L.A.~Dikii, Funct. Anal. Appl. 11 (1977) 93-104. 

\bibitem{gib}J.~Gibbon, P.~Radmore, M.~Tabor and D.~Wood, 
Stud. Appl. Math. 72 (1985) 39-63. 

\bibitem{hone}A.N.W.~Hone, J. Phys. A 30 (1997) 7473-7483. 

\bibitem{hone2}A.N.W.~Hone, {\it Reciprocal links for 
2+1-dimensional extensions of shallow water equations}, 
in preparation. 

\bibitem{pick}C.~Gilson and A.~Pickering, J. Phys. A 28 (1995) 2871-2888. 

\bibitem{krich}I.M.~Krichever, Funct. Anal. Appl. 12 (1978) 59-61. 

\bibitem{newell}A.C.~Newell, {\it Solitons in Mathematics and Physics}, SIAM, 
Philadelphia (1985). 

\bibitem{ohta}Y.~Ohta, J.~Satsuma, D.~Takahashi and T.~Tokihiro, 
Prog. Theor. Phys. (Suppl.) 94 (1988) 210-241. 

\bibitem{okamoto}K.~Okamoto, Physica D 2 (1981) 525-535. 

 \bibitem{satsuma}J.~Satsuma, J. Phys. Soc. Japan 46 (1979) 359-360. 

\bibitem{schiff}J.~Schiff, {\it The Camassa-Holm Equation: A Loop Group Approach}, 
Physica D 121 (1998) 24-43. 

\bibitem{schiff2}J.~Schiff, {\it Integrability of 
Chern-Simons-Higgs Vortex Equations and a Reduction of the 
Self-Dual Yang-Mills Equations to Three Dimensions}, 
in Painlev\a'{e} Transcendents, 393-405, eds. D.~Levi and 
P.~Winternitz, NATO ASI Series B Vol. 278, Plenum Press (1992). 


\bibitem{wtc}J.~Weiss, M.~Tabor and G.J.~Carnevale, J. Math. Phys. 24 (1983) 522-526. 

\bibitem{whit}E.T.~Whittaker and G.N.~Watson, {\it A Course of Modern Analysis} (4th edition), 
Cambridge University Press (1946). 

\end{thebibliography}
\end{document}